\documentclass[aps,prd,groupedaddress,preprintnumbers,nofootinbib,11pt]{revtex4-2}

\usepackage{graphicx,amsmath,amssymb}
\usepackage{xcolor}
\usepackage{ulem}
\usepackage{hyperref}
\usepackage{tikz}
\usepackage{orcidlink}

\hypersetup{colorlinks, linkcolor = [rgb]{0,0.08,0.45}, citecolor = [rgb]{0,0.08,0.45}, urlcolor = [rgb]{0,0.08,0.45}}

\newcommand{\mbf}[1]{\mathbf{#1}}
\newcommand{\req}[1]{(\ref{#1})}
\newcommand{\lb}{\label}
\newcommand{\nn}{\nonumber}
\newcommand{\half}{\textstyle \frac{1}{2}}

\begin{document}

\preprint{SLAC-PUB-17656}

\title{Onset of color transparency in holographic light-front QCD\footnote{Contributed to the MDPI special issue \href{https://www.mdpi.com/journal/physics/special_issues/CTHSRNNCS}{``The Future of Color Transparency, Hadronization and Short-Range Nucleon-Nucleon Correlation Studies”} \vspace{10pt}}}

\author{Stanley~J.~Brodsky\orcidlink{0000-0001-8786-3172}}
\email{sjbth@slac.stanford.edu}
\affiliation{SLAC National Accelerator Laboratory, Stanford University, Stanford, CA 94309, USA}

\author{Guy~F.~de~T\'eramond\orcidlink{0000-0001-6035-7050}}
\email{guy.deteramond@ucr.ac.cr}
\affiliation{Laboratorio de F\'isica Te\'orica y Computacional, Universidad de Costa Rica, 11501 San Jos\'e, Costa Rica}


\begin{abstract}

\vspace{20pt}

The color transparency (CT) of a hadron, propagating with reduced absorption in a nucleus, is a fundamental property of 
QCD (quantum chromodynamics) reflecting its internal structure and effective size when it is produced at high transverse 
momentum,  $Q$. CT has been confirmed in many experiments, such as semi-exclusive hard electroproduction, 
$e A \to e' \pi X$ for mesons produced at $Q^2 > 3 ~ {\rm GeV}^2$.  However, a recent  JLab (Jefferson Laboratory)
measurement for a proton electroproduced in carbon $e\, {\rm C}\to e' p X$, where $X$ stands for the inclusive sum of all produced final states,  fails to observe CT at $Q^2$ up to 14.2 GeV$^2$. In this paper, the onset of CT is determined  by comparing the $Q^2$-dependence of the hadronic cross sections for the initial formation of a small color-singlet configuration using the generalized parton distributions from holographic light-front QCD. A critical dependence on the hadron's  twist, $\tau$, the number of hadron constituents, is found for the onset of CT,  with no significant effects from the nuclear medium. This effect can explain the absence of proton CT in the present kinematic range of the JLab experiment. The proton is predicted to have a “two-stage” color transparency with the onset of CT differing for the spin-conserving (twist-3, $\tau=3$) Dirac form factor with a  higher onset in $Q^2$ for the spin-flip Pauli (twist-4) form factor. In  contrast, the neutron is predicted to have a “one-stage” color transparency with the onset at higher $Q^2$ because of the dominance of its Pauli form factor.  The model also predicts a strong dependence at low energies on the flavor  of the quark current coupling to the  hadron.

\end{abstract}

\maketitle

\newpage


\section{Introduction}

Color transparency (CT) is a unique prediction of quantum chromodynamics QCD, the~theory of the fundamental constituents of hadrons, quarks and gluons. CT represents the ability of a hadron produced at large momentum transfer,  $Q$, in a hard exclusive process to transit a nucleus with reduced absorption. This property reflects the fact that the dynamics of a hadron produced in a hard scattering reaction is dominated by its valence Fock state where its quark constituents have small transverse 
separation,  $a_\perp \propto 1/ Q$, and,~thus, propagate as a small-size \mbox{color-singlet~\cite{Brodsky:1988xz, Brodsky:1994kf, Dutta:2012ii, Jain:1995dd}}. For~example, the~semi-exclusive electroproduction process $e  A \to  e' H X$, where the hadron $H$ is produced with a large transverse momentum opposite to the scattered lepton, is dominated by the hard scattering of its valence quark constituents with small transverse separation of the order of $1/Q$. The~produced hadron, thus, propagates as a small color-singlet, and~the effects of the nuclear environment are suppressed.  The~same principle underlies the theory of hard exclusive reactions~\cite{Lepage:1980fj}, as~well as predictions such as  QCD counting rules for hadronic form factors (FFs) and hard-scattering exclusive cross sections~\cite{Brodsky:1973kr, Brodsky:1974vy, Matveev:ra}.  Here, $X$ stands for  the inclusive sum of all produced final states. Measurements of CT can, thus, distinguish between the predictions  of conventional hadronic and nuclear physics and the onset of quark and gluonic degrees of~freedom.

The reduced nuclear absorption as a function of momentum transfer has been confirmed in many semi-exclusive hard scattering reactions including the electroproduction of mesons, $e A \to e' M X$,  where an isolated meson is produced at high transverse momentum squared of  order $Q^2 > 3 ~ {\rm GeV}^2$. Overviews of the CT measurements are given {in}  Refs.~\cite{Jain:1995dd, Dutta:2012ii, ElFassi:2012hsy}. However, a~recent measurement of the electroproduction of protons by the Hall C Collaboration at 
the Jefferson Laboratory (JLab)~\cite{Bhetuwal:2020jes}, does not observe CT in quasielastic $^{12}{\rm C}(e,e’p)$ for $Q^2$ up to 14.2 GeV$^2$, thus, setting strong constraints for the onset of CT for~baryons.

The  analysis of CT presented in this paper is based on the results from holographic light-front QCD (HLFQCD)~\cite{Brodsky:2014yha, Brodsky:2020ajy}, a~new approach to hadron physics rooted on the gauge/gravity correspondence~\cite{Maldacena:1997re}, LF quantization~\cite{Dirac:1949cp}, superconformal quantum mechanics~\cite{Fubini:1984hf, deTeramond:2014asa, Dosch:2015nwa} and the generalized Veneziano model, including external currents~\cite{Veneziano:1968yb, Ademollo:1969wd, Landshoff:1970ce}. The~framework incorporates features of the hadron 
spectrum, which are not obvious  from the QCD Lagrangian, such as confinement, chiral symmetry breaking and the connection between the baryon and the meson spectrum. The~  HLFQCD  framework also provides nontrivial connections between the dynamics of FFs and quark and gluon distributions by incorporating  Regge behavior at small longitudinal momentum fraction, $x$,  
and the inclusive-exclusive connection at large $x$~\cite{deTeramond:2018ecg, Liu:2019vsn, deTeramond:2021lxc}.

The nonperturbative features of HLFQCD, and,~in particular, the analytic structure of the generalized parton distributions 
(GPDs)~\cite{Mueller:1998fv, Radyushkin:1996nd, Ji:1996ek} , are directly relevant for the analysis of the dynamics underlying 
 CT.  The  effective transverse-impact distance of a hadron as a function of $Q^2$  is found to depend  on hadron's
twist, the~number of quark and gluon constituents of a given Fock state, as~well as the quark current which couples to the hadron. The~hard interaction with the lepton can trigger the initial formation of a small color-singlet configuration which can then propagate with minimal interaction in the nuclear medium. The~dependence on the number of constituents of the hadron's valence Fock state is critical for controlling the onset of color transparency; for example, the~color transparency of a proton or deuteron is highly delayed compared with the observed onset of pion transparency, with~no significant effects from final state inelasticity. The~dependence on twist reflects the fact that the momentum transfer required to bring all of the  constituents of the hadron's valence Fock state to small transverse separation, and,~thus, create a small size color-singlet, grows with the number of constituents. In~fact, this mechanism could explain the sudden drop of CT in quasi-elastic proton scattering $A(p,2p)$ reaction off nuclei~\cite{Carroll:1988rp}, observed precisely at the crossing of the charm threshold~\cite{Brodsky:1987xw}. Different interpretations  are given in Refs.~\cite{Ralston:1988rb, Lee:1992rd}.

\section{Physics of Color Transparency \lb{PCT}}

The connection between GPDs and CT has been the subject of previous theoretical studies~\cite{Burkardt:2003mb, Liuti:2004hd}. In~this paper, this connection is examined within the framework of HLFQCD, which provides analytic expressions to describe the quark and gluon GPDs in the full $x$ and $t = - Q^2$ domains.

\subsection{The effective transverse size of a hadron at large longitudinal momentum fraction $x$}

 The flavor  FF of a hadron can be written in terms of its GPD, $H_q(x,t) \equiv H^q(x, \xi = 0, t) = \rho_q(x,t)$, at zero skewness, $\xi$,  
\begin{align} \lb{Fqf}
F^q(t) & =   \int_0^1 dx \, H^q(x, t) \nn  \\
  & =  \int_0^1 dx \, q(x)  \exp \left[ t \sigma(x)\right],  
\end{align}
where  $q(x)$ is the longitudinal parton distribution function (PDF) and $\sigma(x)$ is the profile function. In~LF quantization, the flavor FF has an exact single-particle representation in impact space (see Appendix \ref{LFFFPD})
\begin{align} \lb{Fxqa}
F^q(q^2) =  \int^1_0 dx  \int d^2 \mbf{a}_\perp e^{i \mbf{a}_\perp \cdot  \, \mbf{q}_\perp} q(x, \mbf{a}_\perp),
\end{align} 
where $\mbf{a}_\perp$ is the $x$-weighted transverse position of the $n-1$ spectator partons,
\begin{align} \lb{aperp}
\mbf a_\perp = \sum_{j=1}^{n-1} x_j \mbf{b}_{\perp j}.
\end{align}
The transverse  coordinate, $\mbf{a}_\perp$, the~transverse-impact parameter, is conjugate to the transverse 
momentum,  $\mbf{q}_\perp$, and $x$  is the longitudinal momentum fraction of the active quark. The~relative transverse variable, $\mbf{b}_{\perp j}$, in~impact space, is the variable conjugate to the relative transverse momentum, $\mbf{k}_{\perp j}$, of particle $j$ with longitudinal momentum fraction  $x_j$. The~index $j$ is summed over the $n-1$ spectator~quarks.

The LF distribution, $q(x, \mbf{a}_\perp)$,  in transverse-impact space is the Fourier transform of the distribution $\rho_q(x, t) = H^q(x,t) = q(x)  
\exp 
\left[ t \sigma(x)\right]$~\cite{Soper:1976jc, Burkardt:2000za, Burkardt:2002hr, 
Brodsky:2006uqa}: 
\begin{align}
q(x, \mbf{a}_\perp) &= \int \frac{d^2 \mbf{q}_\perp}{(2 \pi)^2} \, 
e^{-i \mbf{a}_\perp \cdot \mbf{q}_\perp} \rho_q\left(x, \mbf{q}_\perp \right) \nn \\
&= \frac{1}{4 \pi}  \frac{q(x)}{\sigma(x)}  \exp\left( - \frac{\mbf{a}_\perp^2 }{4 \sigma(x)} \right) .
\end{align}
Let us remark that in Ref.~\cite{Burkardt:2000za,Burkardt:2002hr}, the transverse impact-parameter variable $\mbf{a}_\perp$ is labeled 
$\mbf{b}_\perp$, whereas in 
Ref.~\cite{Brodsky:2006uqa}, 
it is labeled $\vec \eta$.

The distribution function $q(x, \mbf{a}_\perp)$ represents the number density of quarks of flavor $q$ with longitudinal momentum $x$ 
and transverse-impact distance, $\mbf{a}_\perp$, in a given hadron~\cite{Burkardt:2000za,Burkardt:2002hr}. The $x$-dependent transverse-impact distance squared is then given by
\begin{align} \lb{a2x}
\langle \mbf{a}^2_{\perp} (x) \rangle^q &= 
\frac{\int d^2 \mbf{a}_\perp   \mbf{a}_\perp^2  q(x, \mbf{a}_\perp)}{\int d^2 \mbf{a}_\perp  q(x, \mbf{a}_\perp)}  \nn \\
& =  - \frac{1}{\rho_q \left(x, t \right)} \nabla_\mbf{Q}^2 \, \rho_q \left(x,  t \right) \Big  \vert_{t = - Q^2 = 0} \nn \\
&= 4 \,  \sigma(x).
\end{align}

In HLFQCD $\sigma(x)$ is flavor independent and, for hadron twist $\tau$, the~number of 
 hadron constituents in a given Fock component, the  FF has the reparametrization-invariant integral representation,
expressed in terms of Euler’s Beta function~\cite{deTeramond:2018ecg}:
\begin{align}  \lb{FFB}
 F(t)_\tau &=  \frac{1}{N_\tau} B\left(\tau - 1, 1 - \alpha(t)\right) \nn \\   
 &= \frac{1}{N_\tau} \int_0^1 dx\, w'(x) w(x)^{-\alpha(t)} \left[1 - w(x) \right]^{\tau -2} ,
\end{align}
where $\alpha(t) = \alpha(0) + \alpha’ t$ is the Regge trajectory of the vector meson which couples to the quark current in the hadron, and~$N_\tau$ is a normalization factor.  The~trajectory $\alpha(t)$ can be computed  within the superconformal 
 LF holographic framework, and the intercept, $\alpha(0)$, incorporates the quark masses~\cite{deTeramond:2014asa, Dosch:2015nwa}. The~function $w(x)$ is a flavor-independent 
function 
with  $w(0) = 0, \, w(1) = 1$ and   $w'(x) \ge 0$, where the prime defines $x$-derivative. The~profile function $\sigma(x)$ and the PDF 
$q_\tau(x)$ are determined by $w(x)$:
\begin{align} 
\sigma(x)&=\frac{1}{4\lambda}\log\Big(\frac{1}{w(x)}\Big) \lb{sigx} ,  \\ 
q_\tau(x)&=\frac{1}{N_\tau} w'(x)  w(x)^{-\alpha(0)} [1-w(x)]^{\tau-2} \lb{qx} ,
\end{align}
with $\alpha’ = 1 / 4 \lambda$. Boundary conditions  follow from the Regge behavior at $x \to 0$, $w(x) \sim x$,  and~ at $x \to 1$ from the inclusive-exclusive counting rule~\cite{Drell:1969km, Brodsky:1979qm}, $q_\tau(x) \sim (1-x)^{2 \tau - 3}$, which fix $w'(1) = 0 $. These physical conditions, together with the above-given constraints, basically determine the form of $w(x)$.

\begin{figure}[]
\includegraphics[width=8.6cm]{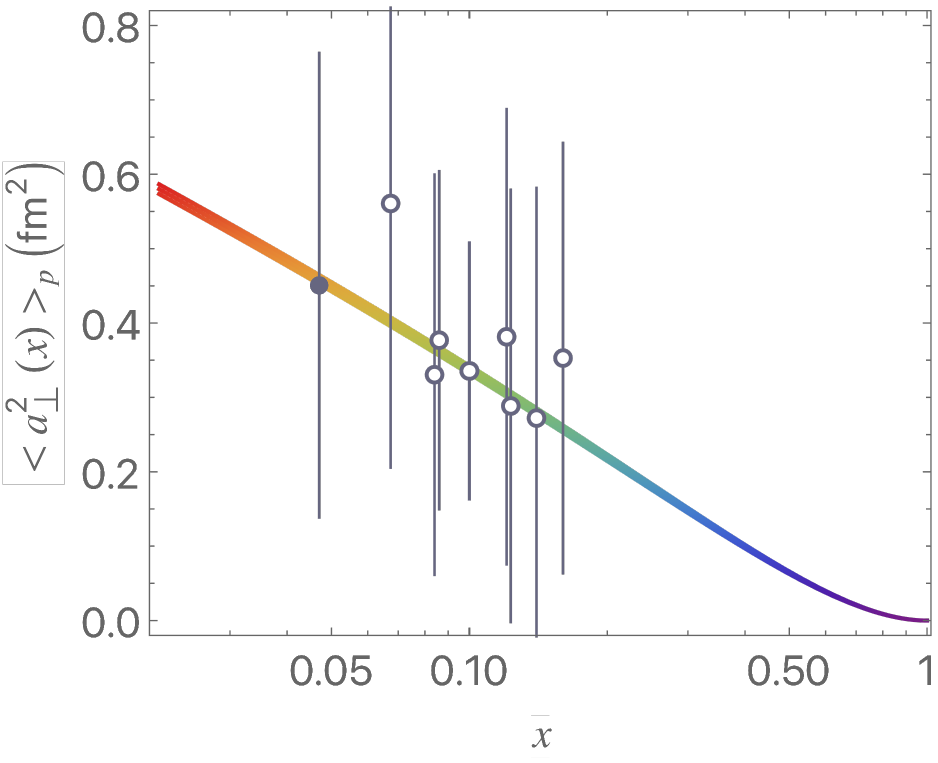}
\includegraphics[width=8.8cm]{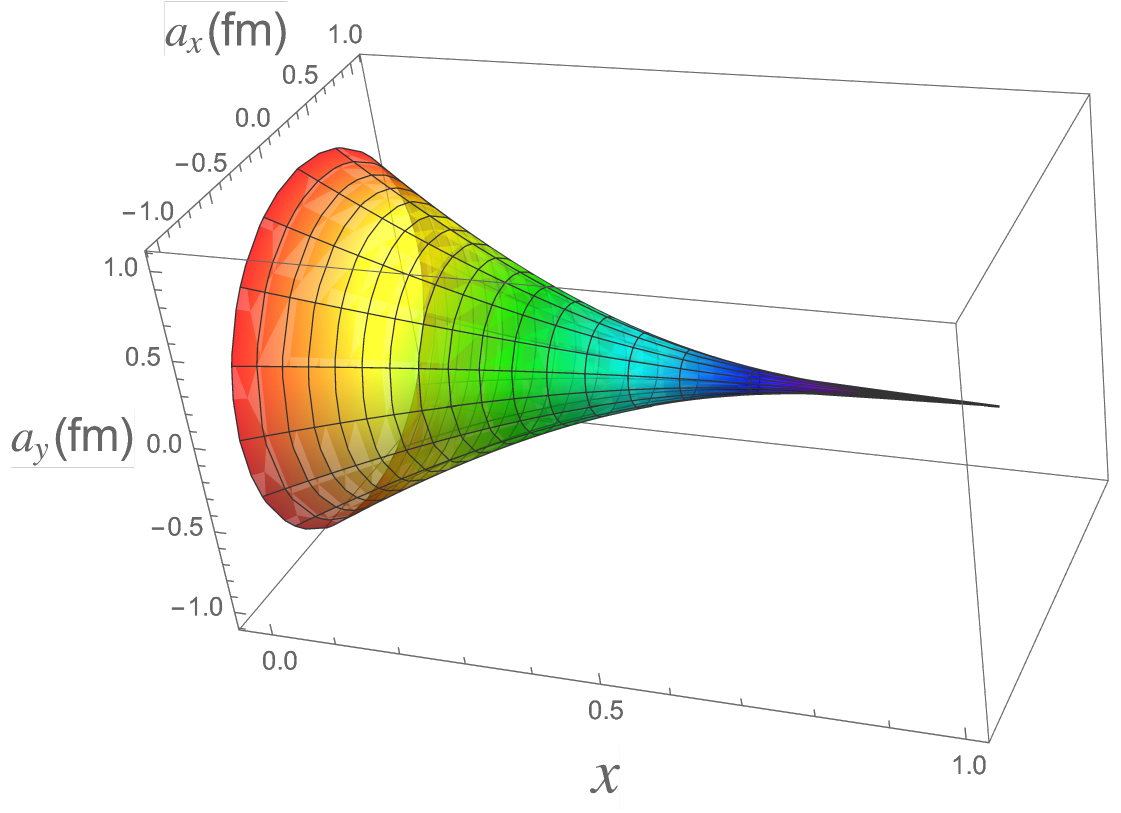}
\caption{\label{ax} The  $x$-dependence of the proton transverse-size, $\langle \mbf{a}^2_{\perp} (x) \rangle_p$, 
determined by the hadron’s profile function, $\sigma(x)$ (Equation~\req{a2x}). {\bf Top:} {Comparison of 
 the results obtained with the data, extracted in  Ref.~\cite{Dupre:2016mai} from the data by CLAS and HERMES experiments. The solid circle represents the HERMES datum and  open circles  the CLAS data. The band
represents the model uncertainty.}  {\bf Bottom:} At large light-front momentum fraction, $x$, and~equivalently, at large values of the momentum transfer squared, $Q^2$, the~transverse size of a hadron behaves as a point-like color-singlet object. This behavior is the origin of color transparency in~nuclei. The rainbow color gradient in the figure represents the transition from the ultraviolet to the infrared domains.}
\end{figure}

Following Ref.~\cite{Dupre:2016mai},  the $x$-dependent transverse-impact distribution of the proton is computed here for the model of the PDFs given in Ref.~\cite{deTeramond:2018ecg}
\begin{align}
u(x) &= \left(2-\frac{r}{3}\right)q_{\tau=3}(x) +\frac{r}{3} \, q_{\tau=4}(x), \label{ux}\\
d(x) &=  \left(1-\frac{2 r}{3}\right) q_{\tau=3}(x) +\frac{2 r}{3} \, q_{\tau=4}(x) \label{dx} ,
\end{align}
ignoring the sea contribution. The~result, including comparison with available data, is shown in Figure~\ref{ax}. The profile function $\sigma(x)$ and the PDF $q_\tau(x)$  are given by Equations \req{sigx} and \req{qx}, with~the result $\langle \mbf{a}^2_{\perp} (x) \rangle_p = 4 \sigma(x)$, independent of the value of $r$ in the model PDFs in Equations \req{ux} and \req{dx}.  The specific form of $w(x)$ is used here as given in Refs.~\cite{deTeramond:2018ecg, Liu:2019vsn,deTeramond:2021lxc},   $w(x) = x^{1-x} e^{-\alpha(1-x)^2}$ with $\alpha =0.5 \pm 0.05$,  and~the precise value of the mass scale, $\kappa \equiv \sqrt{\lambda} =  0.534$ GeV, determined from the Regge trajectories for the $\rho$ and $\phi$ vector mesons~\cite{Sufian:2018cpj}. This value of $\sqrt{\lambda}$ lies within the uncertainty bound of $\sqrt{\lambda} =  0.523 \pm 0.024$ GeV, determined from the full mass spectrum of the light hadrons, including all radial and orbital excited states~\cite{Brodsky:2016yod}. Let us note that similar results have been found in Ref.~\cite{Xu:2021wwj}  for the nucleon and in Ref.~\cite{Adhikari:2021jrh} for pions and kaons.  As~it is shown in Figure~\ref{ax}, at  large $x$, 
the proton converges to its point-like configuration (PLC). 
A similar behavior is predicted at large values of $Q^2$, as~it is expected in a 
very high momentum transfer reaction; this discussed in Section~\ref{sect:larget} below.

For comparison with the results of Section~\ref{sect:larget}, one can also define the $x$-independent transverse-impact distance squared by taking the $x$-average~\cite{Dupre:2016mai},
\begin{align} \lb{aperpq}
\langle \mbf{a}^2_{\perp}\rangle^q   = \frac{1}{N_q} \int dx\, q(x)  \langle \mbf{a}^2_\perp (x) \rangle^q,
\end{align}
and compute, for~example, the~transverse radius of the proton from the charge weighted sum,
$\langle \mbf{a}^2_{\perp} \rangle_p = 2 e_u \langle \mbf{a}^2_\perp\rangle^u + e_d \langle \mbf{a}^2_\perp \rangle^d$. 
Here, $N_q$ denotes the number of valence quarks, $N_u = 2$ and $N_d = 1$.   Ignoring the sea contribution in the proton, Equation~\ref{aperpq} gives the value $\langle \mbf{a}^2_{\perp} (x) \rangle_p = 0.36$ fm compared to the measured value,  $\langle \mbf{a}^2_{\perp} (x) \rangle_p = 0.43 \pm 0.01$~ fm, obtained  from electron-proton scattering~\cite{Dupre:2016mai}.

\subsection{The effective transverse size of a hadron at large momentum transfer $t$ and the onset of color transparency \lb{sect:larget}}

It is shown above that the dependence of the hadron's transverse-size {squared} $\langle \mbf{a}^2_{\perp}(x) \rangle^q$ on the longitudinal momentum fraction $x$ is {flavor independent}  and it is uniquely determined by the hadron profile function $\sigma(x)$: $\langle \mbf{a}^2_{\perp}(x) \rangle^q= 4\ \sigma(x)$. Its behavior in $t = -Q^2$, however, depends on specific properties of the hadron. In~fact, one expects from general considerations that the initial formation of a PLC for a bound state with a large number of constituents---the deuteron for example, with~its large phase space, has a lower probability to fluctuate to a small configuration as compared with a two-particle bound state, say the pion.  Consequently, it presents to the nuclear environment a larger transverse impact area as it traverses through the nucleus, and~it will be slowed down or absorbed with greater probability as compared with a pion projectile with a smaller transverse impact area for the same $Q^2$. The~particle with a larger number of constituents will thus require a  larger $Q^2$ to have the same transparency: the onset of color transparency will be higher in $Q^2$ compared to a hadron with fewer~constituents.

 A similar effect is expected when comparing the relative CT of a nucleon to that of a pion, where the detailed dependence on the individual constituents in the LF wave function (LFWF) is essential.  The~integrand of 
Equation~\req{rhoxq} is in fact a function of the transverse position of the $n-1$ spectator components 
$\mbf{q}_\perp \! \cdot \sum_{j =1}^{n - 1} x_j \mbf{b}_{\perp j} = \mbf{q}_\perp \cdot \mbf{a}_\perp$ (see Equation~\req{aperp}):  It corresponds to a change of transverse momentum $x_j \mbf b_{\perp j}$ for each of the $n-1$ spectator particles. This dependence is crucial for determining the relative CT of different hadrons since it measures the transverse size of a scattered hadron in a given Fock~state.

The transverse impact surface dependence on the momentum transfer, $t= - Q^2$, is computed from the expectation value of the profile function $\sigma(x)$:
\begin{align} \lb{a2tF}
\langle 4 \sigma(t) \rangle^q  & =  
\frac{\int dx \, 4 \sigma(x) \rho_q(x, t)}{\int d x \rho_q(x,  t)}  \nn \\
& =   \frac{4}{F^q (t)} \frac{d}{d t} F^q (t),
\end{align}
with  the GPD $ H^q(x,t) = \rho_q(x, t) = q(x) \exp\left[t \sigma(x)\right]$, integrated over the longitudinal variable $x$ of the active quark. From Equation~\req{a2x} it follows that  $\langle 4 \sigma(t) \rangle^q$ at $t =0$ is precisely the $x$-independent transverse-impact distance squared  \req{aperpq}, namely,  $ \langle {a}_\perp^2 \rangle^q = \langle 4 \sigma(t = 0) \rangle^q$. The~transverse impact surface, $\langle 4\sigma(t)  \rangle$, thus, measures the slope of the hadron  FF at any value of the momentum transfer, $ Q^2 = - t > 0$:  the transverse distance naturally evolves from a small color-singlet configuration at large $Q^2$ to the  actual equilibrium size at $Q^2 = 0$ compatible with the usual definition of the hadron radius:  $\langle \mbf{a}_\perp^2 \rangle =  \frac{2}{3}\langle r^2\rangle $.

In the nonperturbative framework, presented here, the GPDs incorporate the far-off-shell components of the LFWF, which controls the behavior of the  FF at large $Q^2$, and~the power counting rules from the inclusive-exclusive connection~\cite{deTeramond:2018ecg}.  For a given twist-$\tau$ Fock 
component in the hadron wave function one finds: 
\begin{align} \lb{a2t}
\langle 4 \sigma(t) \rangle_\tau   =  
\frac{1}{\lambda} \left[ \psi\left(\tau - \alpha(t)\right) - \psi(1 - \alpha(t) \right],
\end{align}
a result which follows directly from the expression of the FF, given in Equation~\req{FFB}, since $ B(u,v)^{-1} \partial_v B(u,v) = \left( \psi(v) - \psi(u + v)\right)$, with~$\psi(z)$ the digamma function, $\psi(z) = \Gamma(z)^{-1}  {d}\Gamma(z)/{d z} $. For~integer twist, $\tau = N$,  the recurrence relation for the digamma function, $\psi(z + 1) - \psi(z) = {1}/{z}$, 
can be used to obtain
\begin{align} \lb{a2N}
\langle 4 \sigma(t) \rangle_\tau   =  \frac{1}{\lambda} \sum_{j = 1}^{\tau-1} \frac{1}{j - \alpha(t)},
\end{align}
an expression reminiscent of the classical Regge pole structure of a scattering amplitude. In~contrast to the dependence of the transverse impact area as a function of $x$, given by Equation~\req{a2x}, Equation~\req{a2N} depends explicitly on the hadron's twist, $\tau$, and the properties of the specific quark current, which couples with the active quark in the hadron, characterized by the hadron's Regge trajectory,
$\alpha(t)$.  For~large values of $ t = - Q^2$, Equation~\ref{a2N} leads to
\begin{align} \lb{a2tasy}
\langle \sigma(t) \rangle_\tau   \to  \frac{(\tau - 1)}{Q^2},
\end{align}
which shows that, as~expected, the~hadronic size decreases with $Q^2$. Physically, the~$Q^2$ required to contract all of the valence constituents of a hadron to a color-singlet domain of a given transverse size, grows as the number of the spectator constituents as $\tau-1$.~Let us note that the applied procedure to compute the $Q^2$-dependence of the effective transverse size differs from the procedure of  Ref.~\cite{Burkardt:2003mb}, given instead in terms of the relative impact variables. The procedure of Ref.~\cite{Burkardt:2003mb} leads to a rising transverse-impact surface with increasing $Q^2$ and decreasing twist, in~contradiction with the results, presented here, and 
the observations~\cite{Miller:2022kxt}.

\subsection{Inelastic corrections to the onset of color transparency \lb{FSI}} 

 One has to note  that additional effects for the onset of CT from the interaction of the scattered hadron, as~it propagates through nuclear matter, do not modify the onset of CT, provided that the momentum transfer $Q^2$ is relatively large. To~show this, the inelastic scattering cross section  is evaluated assuming that the effects from the expansion of the small scattered hadron are not significant during the collision time as it transits out of the 
nucleus~\cite{Farrar:1988me, Caplow-Munro:2021xwi}. In~this approximation, one can use the inelastic cross section formula for the scattering of a small dipole of transverse size, $d_\perp$, off a nucleon in the two-gluon exchange approximation~\cite{Blaettel:1993rd, Frankfurt:1993it, Frankfurt:1996ri}: 
\begin{align} \lb{da}
\sigma^{\rm inel}_\tau(Q^2,x) = \frac{4 \pi^2}{3} \alpha_s(Q^2) \langle \sigma_\tau(Q^2)\rangle x  G_N(x,Q^2),
\end{align}
where $G_N(x,t)$ is the gluon distribution in the nucleon, which underlies the inelastic cross section for the scattered hadron, and $x \sim Q^2 /s$, where $s$ is the center of mass energy squared of the colliding particles. The intrinsic gluon distribution in {the pion} and nucleons was studied in Ref.~\cite{deTeramond:2021lxc} in the full range of $x$ and $t$.   

Deriving Equation \req{da}, it is assumed  that the size of the small color dipole $d^2_\perp$ is small relative to the equilibrium size of the proton, namely, 
$d_\perp^2 \ll \langle \mbf{a}_\perp^2 \rangle$, with~${\langle \mbf{a}_\perp^2 \rangle } = \langle 4 \sigma(t =0)\rangle$. 
 One, thus, has from Equations
\req{a2t} and \req{a2tasy}:
\begin{align}
    d_\perp^2(Q^2) \simeq  \langle 4 \sigma(Q^2)\rangle \sim \frac{4 (\tau - 1)}{Q^2},
\end{align}
for relatively large $Q^2$, and~\begin{align}
\frac{\sigma^{\rm inel}_{\tau_A}(Q^2,x)}{\sigma^{\rm inel}_{\tau_B}(Q^2,x)} 
 \simeq  \frac{\langle \sigma_{\tau_A}(Q^2)\rangle}{\langle 
\sigma_{\tau_B}(Q^2)\rangle},
\end{align}
with no significant {twist-dependent} effects from final state inelasticity;  here,  $\tau_A$ and $\tau_B$ denote the twists of the hadrons $H_A$ and $H_B$ electroproduced, respectively, in the reaction $e A \to e’ H X$ at large $Q^2$. Therefore, to a first approximation, the onset of CT for the scattered hadron in nuclei only depends on the ratio of the transverse impact surface at large enough $Q^2$, thus, on the relative~twist.

\subsection{The Feynman mechanism}

It has been argued~\cite{Caplow-Munro:2021xwi} that the ``Feynman mechanism” -- where the behavior of the struck quark at large longitudinal momentum $x \to 1$ in the proton LF wave function is assumed to play a key role for hard exclusive hadronic processes -- does not predict color transparency. This process takes place when an active quark, carrying a significant fraction of the nucleon’s momentum, is {turned} back by the incident virtual photon. In~this case, the spectator quarks are “wee” partons and would not shrink to a small transverse distance; thus color transparency would not be observed~\cite{Caplow-Munro:2021xwi}.

We stress that the CT mechanism is applicable to hard exclusive scattering events which are controlled by valence Fock states:~for example, in $ep \to e’ p’$, where the entire proton changes direction, and~one has, as~well, to~take into account the transverse dependence of the LFWF. Each spectator quark $j$  in the final-state LFWF has momentum, $x_j \mbf{q}$, and~the transverse  impact distance shrinks to small separation $a_\perp \propto 1/ Q$ as shown above; therefore CT does occur at large $Q^2$ in contrast from the deep inelastic scattering mechanisms where only the struck quark  at large $x$  flips its direction, thus avoiding the formation of a PLC at any $Q^2$.

\section{Discussion of the Results \lb{DoR}}

Figure~\ref{sigQ2} shows that the gap in the effective transverse impact surface, $ \langle 4 \sigma(t) \rangle$, for different twist (the number of constituents of a given Fock  component) is most significant at intermediate energies and for low twist values, particularly between twist-2  ($\tau=2$) and twist-3. For~example, the~effective transverse impact surface for twist-2 at 4 GeV$^2$ is similar to that of twist-3 at 14 GeV$^2$.  Likewise, the~impact surface at 4 GeV$^{2}$ for twist-2 is similar to that of twist-4 at 22 GeV$^2$, thus, implying an important delay in the onset of CT at intermediate energies in terms of the number of quark constituents. This is particularly relevant for the proton since it contains, at approximately equal probability, both twist-3, but~also twist-4, in~its LFWF.  The~twist-4 contribution is required in order to generate the proton's anomalous magnetic moment and
Pauli FF. This predicts a larger value of $Q^2$ in order to produce color transparency for the proton,  consistent  with the recent JLab electroproduction measurement~\cite{Bhetuwal:2020jes}.

\begin{figure}[ht]
\includegraphics[width=8.6cm]{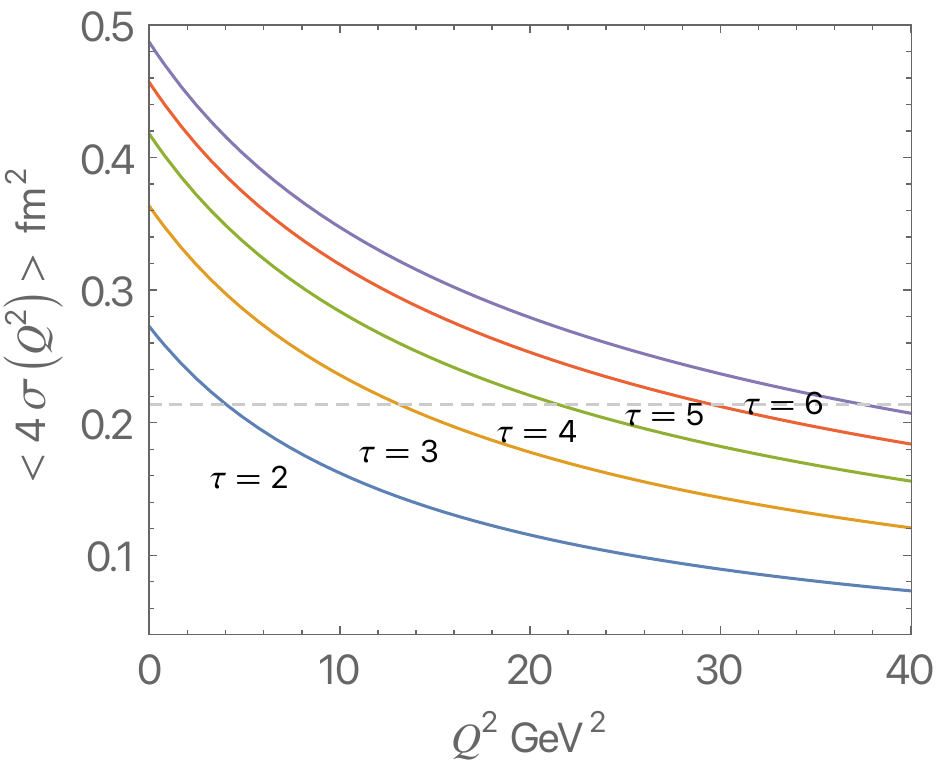}
\caption{The transverse impact area $ \langle 4 \, \sigma \left( t \right) \rangle$ as a function of $Q^2$ and the number of constituents  (twist), $\tau$, implies a significant delay in the onset of color transparency at intermediate energies for $\tau > 2$. 
The~dashed line indicates the characteristic transverse size, required for the onset of color~transparency. \lb{sigQ2}}
\end{figure}

{\subsection{Two-Stage Color Transparency in the~Nucleon}

The initial small hadron configuration in lepton-proton elastic scattering at large $Q^2$ leads to spin-flip and non spin-nonflip transitions. They correspond to a distinctive lepton's angular distribution, specified by the Rosenbluth differential cross-section in terms of the Dirac and Pauli  FFs. Here, the specific distributions \req{ux} and \req{dx} from Ref.~\cite{deTeramond:2018ecg} 
are used to obtain the Dirac FFs for proton and neutron in the approximation where sea components are neglected:
\begin{align}
 F_1^p(Q^2) &= F_3(Q^2), \lb{F1p}\\
 F_1^n(Q^2) &= - \half \left(F_3(Q^2) - F_4(Q^2)\right) , 
\lb{F1n}
 \end{align}
$F_{\tau}$, defined by Equation~\ref{FFB}. 

The result for the proton is independent of the value of $r$, whereas for the neutron it corresponds to $r =  {3}/{2}$ from the anti-de-Sitter (AdS) normalization constraints. The~leading twist-4 result for the Pauli FF is computed from the overlap of LF orbital-angular-momenta $L = 0$ and $L = 1$ components~\cite{Brodsky:2014yha}:
\begin{align}
    F_2^{p,n}(Q^2) = \chi_{p,n} F_3(Q^2),  \lb{F2pn}
\end{align}
where $\chi$ is the nucleon anomalous magnetic moment. In~Equation \req{F2pn}, higher-twist sea contributions are neglected~\cite{Sufian:2016hwn}. Let us note that the Dirac neutron FF \req{F1n} has equal but~opposite sign,  twist-3 and twist-4 components, which leads to a strong cancellation of the Dirac~FF.

The inelastic process, as~the nucleon travels in the nuclear medium, depends on the nucleon's transverse impact surface. For~the proton, the~charge averaged density gives a sensible measure of its impact transverse size. For~the neutron, however, it gives a very small effective size. One can use instead the isoscalar averaged distribution $\frac{1}{3} \left[u(x) + d(x)\right]$
to compute the effective dependence of the nucleon transverse impact size. The~isoscalar component corresponds to the physics of Pomeron exchange and it is also a measurable quantity related to the isospin $I = 0$ nucleon FF combination:  $F^p + F^n$. This definition leads to the  same effective  transverse impact surface for the proton and the neutron and is consistent with AdS baryon 
normalization of equal $L = 0$ and $L =1$ probability~\cite{Brodsky:2014yha}. The~onset of CT from spin-flip transitions is at 22 GeV$^2$ since the Pauli FF is leading twist 4. To~recapitulate,} the proton is predicted to have a “two-stage” color transparency starting  at $Q^2>14~{\rm GeV}^2$ 
for the  twist-3 valence Fock state with  $L=0$ and at $Q^2 > 22~{\rm GeV}^2$ for the full onset of CT  for proton's $L = 0$ and  $L=1$ twist-4 components. {In contrast, the~neutron only presents a “one-stage” color transparency onset at $Q^2 > 22~{\rm GeV}^2$ due to its Pauli  FF.

One can make a more quantitative estimate for the onset of color transparency for the proton vs. the pion by using existing data for pion electroproduction~\cite{Clasie:2007aa,Qian:2009aa}. In~the case of $e  ^{12}{\rm C} \to e \pi X$, measurements indicate that the transparency ratio, $T$, for pions rises approximately  10\%  as  $Q^2$ increases from 3 to 4~${\rm GeV}^2$; see Figure~14 Ref.~\cite{Dutta:2012ii}.  As~shown in Figure~\ref{sigQ2}, the~observation of CT in carbon in this range of $Q^2$ requires that the transverse surface $\langle 4 \sigma(Q^2)\rangle$ of the pion ($\tau =2$)  has contracted approximately to $0.22~ {\rm fm}^2$, indicated by the dashed line in Figure~\ref{sigQ2}.  It is also clear from Figure~\ref{sigQ2} that one  will obtain a similar size  \mbox{$\langle 4 \sigma \rangle \simeq 0.22~{\rm fm}^2$} for the $\tau=3$ proton at  $Q^2 \simeq 14~{\rm GeV}^2 $, and~much higher \mbox{$Q^2 \simeq 22~ {\rm GeV}^2$} for the $\tau =4$ proton {component}. However, since the proton is produced with approximately equal $\tau=3$ and $\tau=4$ components, one predicts only a small 5\% increase of the transparency, i.e.,~from $T= 0.56$ to $T=0.59$, for~$e ^{12}{\rm C} \to e p X$ at $Q^2 = 14~{\rm GeV}^2$.  This small variation in transparency is compatible with the recent measurement of proton electroproduction in carbon~\cite{Bhetuwal:2020jes}. The~effect of CT   
 is, therefore, expected to be larger in a heavier nuclear~target.

CT for the production of an intact deuteron nucleus in \mbox{$e A \to d + X_{(A-2)}$}  quasi-exclusive reactions should be observed at $Q^2 \ge 40~{\rm GeV}^2$. This can be tested in $e d \to e d $ elastic scattering in a nuclear~background.

\begin{figure}[]
\begin{center}
\includegraphics[width=8.6cm]{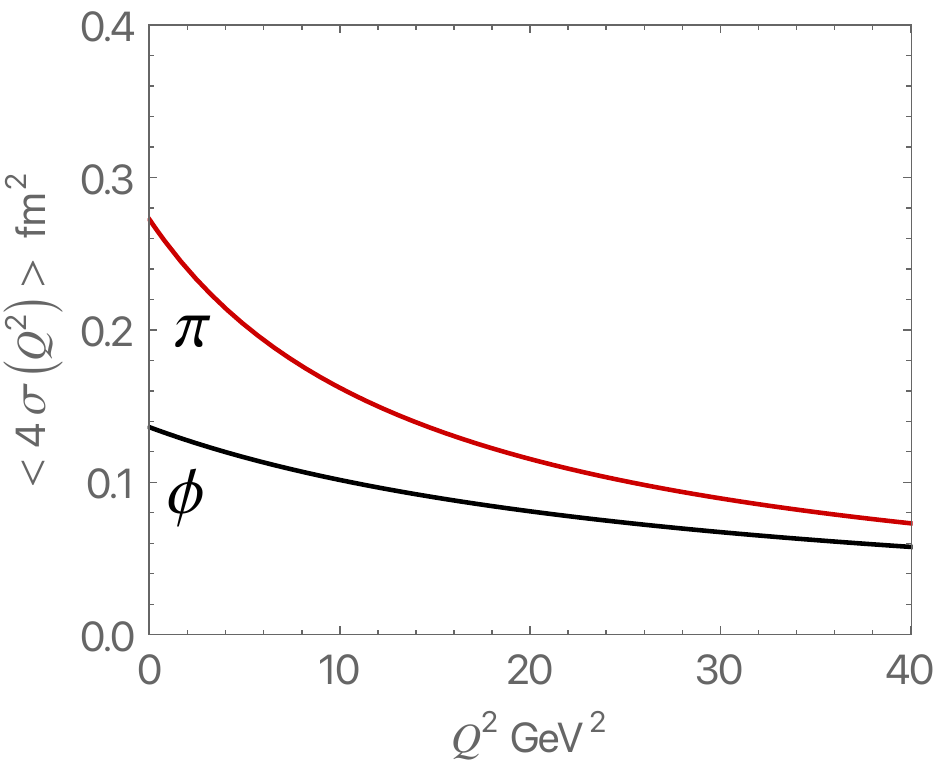}\\
\vspace{-192pt}\hspace{86pt}\includegraphics[width=4.8cm]{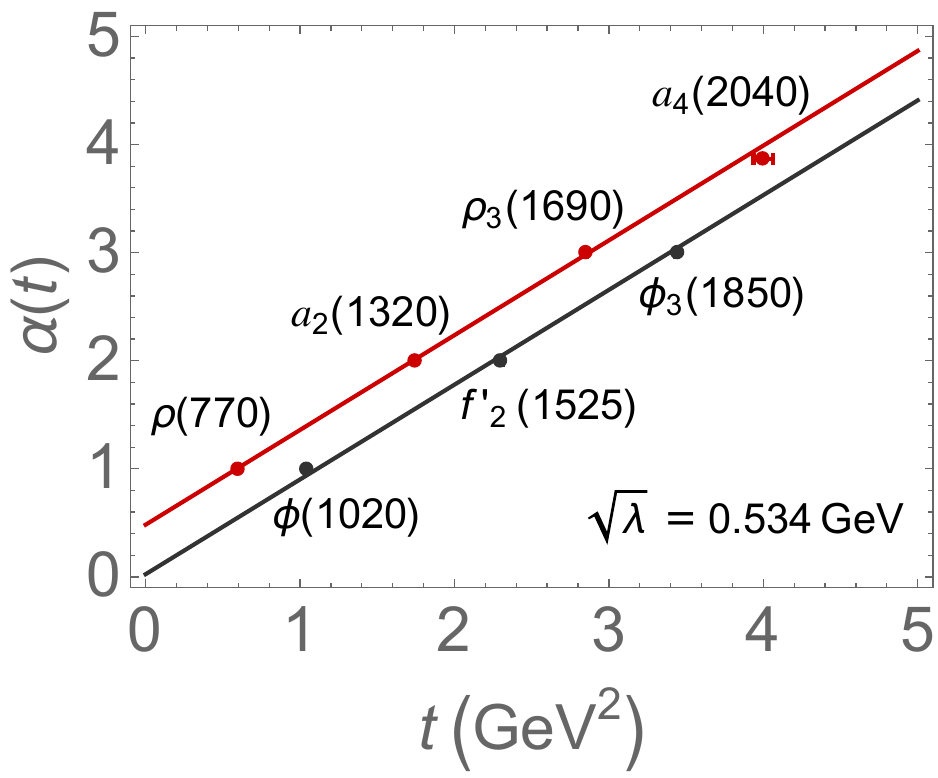}
\end{center}
\vspace{80pt}
\caption{Comparison of the transverse impact surface,  $ \langle 4 \, \sigma \left( t \right) \rangle$, as a function of $Q^2 = -t$ for the 
$\phi$ meson (black) and pion (red). The~small transverse size and the slow fall-off for the $s \bar s$ vector meson reflects the heavier mass of the strange quark. The inset from Ref.~\cite{Sufian:2016hwn}  represents the corresponding linear Regge trajectories for the $\rho$ and $\phi$ vector mesons in Equation~\req{a2N} {{for $\sqrt{\lambda} = 0.534$ GeV}, $\alpha_\rho(0)$ = 0.483, $\alpha_\phi(0)$ = 0.025.} \label{A3Q2tw}}
\end{figure}

\subsection{Effect of Quark Flavor on the Onset of Color~Transparency}

Another remarkable prediction of  the result \req{a2N} is the strong dependence at low energies of the specific quark current coupling to a hadron. To~illustrate this property, Figure~\ref{A3Q2tw} compares $Q^2$-dependences of the transverse impact area, $\langle 4 \sigma(Q^2)\rangle$, for the $\phi$ meson and for the pion. The~small transverse size and its slow fall-off for the $s \bar s$ vector meson reflects the heavier mass of the strange quark. One also sees that $e A \to e \phi X$ electroproduction will show color transparency even at small $Q^2$ and can provide an important calibration tool for transparency~measurements.

\section{Conclusions and Outlook}

Color transparency (CT) in one of the most striking properties of quantum chromodynamics (QCD) phenomenology~\cite{Brodsky:1988xz}:  CT refers to the reduced absorption of a hadron as the latter propagates through nuclear matter if produced at high transverse momentum in a hard exclusive process.  The~nuclear absorption reflects the size of the color dipole moment of the propagating hadron, i.e.,~ the separation between its colored~constituents.

The onset of CT predicted in this paper, is based on the nonperturbative analytic structure of the hadron generalized parton distributions (GPDs), obtained in the framework of holographic  light-front (LF) QCD and valid in the full domain of the kinematic variables. The~GPDs incorporate the Regge behavior at small values of the longitudinal light front momentum fraction,  $x$, and~more relevant for the present paper,  the~inclusive-exclusive connection at large $x$.

A key quantity, which measures the transverse effective size of a scattered hadron in a given Fock state is the impact parameter transverse distance, $\mbf{a}_\perp$, the~$x$-weighted transverse position coordinate of the $n-1$ spectators, Equation \req{aperp}, which, as~shown in  Appendix~\ref{LFFFPD}, is the variable conjugate to the transverse momentum, $\mbf{q}_\perp$. As~such,  $\mbf{a}_\perp$ is the relevant variable which controls the effective transverse distance of the hadron at large momentum transverse squared, $Q^2$: It leads to an effective transverse surface which has the expected scaling properties and a crucial dependence on twist at high $Q^2$. The transverse distance, $\mbf{a}_\perp$, also depends on the flavor of the quark current, which couples to a given~hadron.

The hadronic size decreases inversely to  $Q^2$, and increases with the hadron's twist, $\tau$. This corresponds physically to the fact that the momentum transfer, required to produce a compact hadronic Fock state, increases with the number of its fundamental constituents. An~essential consequence is that at a given momentum transfer, the~effective transverse size for mesons with leading twist $\tau =2$ is smaller than for baryons with $\tau=3$ and $4$, corresponding to Fock states with the orbital LF angular momenta, $ L=0 $ and $ L=1 $, respectively.  Actually,  the~proton is predicted to have a ``two-stage''  CT  onset with $Q^2>14~{\rm GeV}^2$ for the twist-3 Fock state with orbital angular momentum $ L=0$, and~at $Q^2 > 22~{\rm GeV}^2$ for the later full onset of CT for its combined $L = 0$ and $L=1$ twist-4 components. {In contrast, the~neutron is predicted to have a ``one-stage'' color transparency onset at $Q^2 > 22~{\rm GeV}^2$ because of the dominance of its Pauli  FF.

CT is predicted to occur at a significantly  higher momentum transfer $Q^2$ for baryons  ($Q^2 > 14~{\rm GeV}^2$ {for the proton and $Q^2 > 22~{\rm GeV}^2$ for the neutron) as compared with} mesons $(Q^2 > 4~{\rm GeV}^2)$.  Note that, it was already proposed in Ref.~\cite{Blaettel:1993rd}, by~comparing with nucleon-nucleon scattering, that CT effects 
should be more pronounced for a meson beam. The predictions found here for the onset of CT are consistent with the 
findings  at Jefferson Laboratory (JLab), which have confirmed CT presence  for the $\pi$ and $\rho$ mesons~\cite{Dutta:2012ii, ElFassi:2012hsy} and the absence of CT for protons. These existing measurements are, however, limited to values below the range of $Q^2$ where the onset of proton CT is predicted to~occur.

\section*{Acknowledgements}

 We thank Jennifer Rittenhouse West for helpful discussions and Marc Vanderhaeghen for providing the data for the proton transverse size from Ref.~\cite{Dupre:2016mai}.  This work is supported by the Department of Energy, Contract DE--AC02--76SF00515.

\appendix
\section{Form factors and parton distributions in light-front QCD \lb{LFFFPD}}

The light-front (LF) formalism provides an exact representation of current  matrix elements in terms of the overlap 
of frame-independent light-front wave functions (LFWFs) in a LF Fock basis expansion with components $\psi_n(x_i, \mbf{k}_{\perp i},\lambda_i)$, where the internal partonic coordinates, the~longitudinal momentum fraction, $x_i$, and the transverse momentum, $\mbf{k}_{\perp i}$, obey the momentum conservation sum rules $\sum_{i=1}^n x_i =1$, and~$\sum_{i=1}^n \mbf{k}_{\perp i}=0$. The~LFWFs also depend on  $\lambda_i$, the~projection of the constituent's spin along the $z$ direction.

The Drell-Yan-West (DYW) relation~\cite{Drell:1969km, West:1970av} provides a rigorous representation of the  electromagnetic 
 form factors (FFs) of hadrons in terms of the overlap of their LFWFs in the momentum space:
 \begin{multline} \lb{DYW}
F(q^2) =   \sum_n  \prod_{i=1}^n \int 
dx_i \!  \int \frac{d^2 \mbf{k}_{\perp i}}{2 (2\pi)^3} \, 16 \pi^3 \,
\delta \Big(1 - \sum_{j=1}^n x_j\Big) \, \delta^{(2)} \negthinspace\Big(\sum_{j=1}^n\mbf{k}_{\perp j}\Big) \\
\sum_j e_j \psi^*_n (x_i, \mbf{k}'_{\perp i},\lambda_i) 
 \psi_n (x_i, \mbf{k}_{\perp i},\lambda_i),  \
\end{multline}
where  the variables of the LF Fock components in the final state are given by $\mbf{k}'_{\perp i} = \mbf{k}_{\perp i} + (1 - x_i)\, \mbf{q}_\perp $ for a struck  constituent quark and $\mbf{k}'_{\perp i} = \mbf{k}_{\perp i} - x_i \, \mbf{q}_\perp$ for each spectator. 
Equation~\ref{DYW}  represents an exact formula if the sum is taken over all Fock states, $n$.

The DYW expression for the  FF can be expressed in the impact space  by Fourier transforming Equation \req{DYW} in momentum space to impact transverse space~\cite{Soper:1976jc}.  This is a convenient form to obtain the impact dependent representation of 
the generalized parton distributions (GPDs)~\cite{Burkardt:2000za,Burkardt:2002hr}, but~also for the holographic mapping of 
anti-de-Sitter (AdS) results to LF physics, since the DYW FF can be expressed in terms of the product of LFWFS with identical variables~\cite{Brodsky:2006uqa}.  To~this purpose,  Equation \req{DYW} is rewritten in terms of  $n\! - \! 1$ independent transverse impact variables, $\mbf{b}_{\perp j}$, $j = 1,2,\dots,n-1$, conjugate to the relative transverse momentum coordinate, ${\mbf{k}_{\perp j}}$, and~label by $n$  the active charged parton which interacts with the  current. Using the Fourier expansion,
\begin{align} \label{eq:LFWFb}
\psi_n(x_j, \mathbf{k}_{\perp j}) =  (4 \pi)^{(n-1)/2}
\prod_{j=1}^{n-1}\int d^2 \mbf{b}_{\perp j}
\exp{\Big(i \sum_{k=1}^{n-1} \mathbf{b}_{\perp k} \! \cdot \mbf{k}_{\perp k}\Big)} \,
{\psi}_n(x_j, \mathbf{b}_{\perp j}),
\end{align}
 one finds~\cite{Soper:1976jc, Brodsky:2006uqa}:
\begin{align}  \lb{FFb} 
F(q^2) =  \sum_n  \prod_{j=1}^{n-1}\int d x_j  \! \int d^2 \mbf{b}_{\perp j} 
\exp \! {\Big(i \mbf{q}_\perp \! \cdot \sum_{j=1}^{n-1} x_j \mbf{b}_{\perp j}\Big)} \left\vert  \psi_n(x_j, \mbf{b}_{\perp j})\right\vert^2,
\end{align}
corresponding to a change of transverse momentum, $x_j \mbf{q}_\perp$, for each of the $n\! - \!1$ spectators. The internal parton variables, the~longitudinal momentum fraction $x_i$ and the transverse impact coordinates $\mbf{b}_{\perp i}$ obey the sum rule: 
 $\sum_{i=1}^n x_i = 1$ and}} $\sum_{i=1}^n \mbf{b}_{\perp i} = 0$.

The FF in LF  quantization has an exact representation in terms of a single particle  
density~\cite{Soper:1976jc, Brodsky:2006uqa}:
\begin{align} \lb{Fxq}
F(q^2) = \int_0^1 dx ~\rho(x, \mbf{q}_\perp),
\end{align}
where $\rho(x, \mbf{q}_\perp)$ is given by
\begin{multline} \lb{rhoxq}
\rho(x, \mbf{q}_\perp) = \sum_n \prod_{j=1}^{n-1} 
\int dx_j \int d^2 \mbf{b}_{\perp j} \, \delta \Big(1-x - \sum_{j=1}^{n-1} x_j\Big) 
\exp{\Big(i \mbf{q}_\perp \! \cdot \sum_{j=1}^{n-1} x_j \mbf{b}_{\perp j}\Big)}
\left\vert  \psi_n(x_j, \mbf{b}_{\perp j})\right\vert^2.
\end{multline}
The integration in Equation \req{rhoxq} is over the coordinates of the $n-1$ spectator partons, and~ $x = x_n$ is the coordinate of the active charged quark.

One can also express the FF \req{Fxq} in terms of a single-particle transverse distribution, $q(x,\mbf{a}_\perp)$, in the transverse-impact parameter space~\cite{Soper:1976jc}:
\begin{align} \label{FFaq}
F(q^2) =\int^1_0 dx  \int d^2 \mbf{a}_\perp e^{i \mbf{a}_\perp \cdot  \, \mbf{q}_\perp} q(x, \mbf{a}_\perp),
\end{align}
where
$\mbf{a}_\perp = \sum^{n-1}_{j=1} x_j \mbf{b}_{\perp j}$
is the $x$-weighted transverse position coordinate of the $n\! -\! 1$ spectators.  From Equation \req{rhoxq}, 
 one obtains  the corresponding transverse density:
\begin{align} \label{rhoxa}
q(x,\mbf{a}_\perp)   
&= \int \frac{d^2 \mbf{q}_\perp}{(2 \pi)^2}  
e^{-i \mbf{a}_\perp \cdot \mbf{q}_\perp} \rho(x, \mbf{q}_\perp)  \\ 
&= \sum_n \prod_{j=1}^{n-1} \int dx_j \int d^2\mbf{b}_{\perp j} 
\,\delta \Big(1-x-\sum_{j=1}^{n-1} x_j\Big)  \,
 \delta^{(2)}\Big(\sum_{j=1}^{n-1} x_j \mbf{b}_{\perp j} - \mbf{a}_\perp\Big)
\left\vert \psi_n(x_j, \mbf{b}_{\perp j}) \right\vert^2. \nn
\end{align}
The procedure is valid for any Fock state $n,$ and, thus, the results can be summed over $n$ to obtain an exact representation of the impact-parameter-dependent  parton distribution, introduced in Ref.~\cite{Burkardt:2000za,Burkardt:2002hr}.  Then, Equation~\ref{rhoxa} gives the probability to find a quark with longitudinal light front momentum fraction $x$ at a transverse distance $\mbf{a}_\perp$.

One  can also compute the charge distribution of a hadron in the LF transverse plane $\rho(\mbf{a}_\perp)$ by integrating 
Equation \req{rhoxa}. Using Equations \req{Fxq} and \req{rhoxa}, one finds: 
\begin{align} \lb{rhoa}
\rho(\mbf{a}_\perp)  &\equiv  \int _0^1 dx \, q(x,\mbf{a}_\perp) \nn\\
   &= \int \frac{d^2 \mbf{q}}{(2 \pi)^2} \, e^{-i \mbf{a}_\perp \cdot \mbf{q}_\perp}  F(q^2) \nn \\
   &= \int \frac{q dq}{2 \pi}  J_0(a q) F(q^2),
\end{align}
with $q^2 = - Q^2 = t$. Note that Equation  \req{rhoa} matches the expression, obtained  in Refs.~\cite{Miller:2007uy, Miller:2010tz}, for the LF transverse charge density provided one identifies $a_\perp$ with $b_\perp$ here.


\end{document}